\documentclass{aa}
\usepackage{graphicx}
\usepackage{natbib}
\usepackage{amssymb}
\usepackage{relsize}
\usepackage{txfonts}
\usepackage{rotate}
\usepackage{epsf}
\newcommand{\D}{\partial}
\newcommand{\DD}{\frac}

\newcommand{\thet }{\theta}

\newcommand{\beq}{\begin{equation}}
\newcommand{\eeq}{\end{equation}}
\newcommand{\ben}{\begin{enumerate}}
\newcommand{\een}{\end{enumerate}}
\newcommand{\bit}{\begin{itemize}}
\newcommand{\eit}{\end{itemize}}
\newcommand{\barr}{\begin{array}}
\newcommand{\earr}{\end{array}}
\newcommand{\mm }{\mathrm}

\newcommand{\DroI}{{\overline{\rm D}}}

\begin{document}
\title{ Angular momentum transport during X-ray bursts on neutron stars:
a numerical general relativistic hydrodynamical study}

\titlerunning{Low-Mach number flows}

\author{A.~Hujeirat${}^{1}$
        \and F.-K. Thielemann${}^{2}$}

\authorrunning{A.~Hujeirat et al.}
\institute{
ZAH, Landessternwarte Heidelberg-K\"onigstuhl,\\
Universit\"at Heidelberg, 69120 Heidelberg, Germany \and Departement Physik, Universit\"at Basel, Switzerland }

\date{Received ... / Accepted ...}

\offprints{A. Hujeirat,  \email{AHujeirat@lsw.uni-heidelberg.de}}

\abstract
{}
  {The distribution of angular momentum of the matter during X-ray bursts on neutron stars
is studied by means of 3D axi-symmetric general relativistic hydrodynamics.}
{The set of fully general relativistic Navier-Stokes equations is solved implicitly using the
implicit solver GR-I-RMHD in combination with a third order spatial and second order temporal
advection scheme.
The viscous operators are formulated using a Kerr-like metric in the fixed background
of a slowly rotating neutron star whose radius coincides with  the corresponding
last stable orbit.
The importance of these operators and their possible simplifications are discussed as well.
To verify the consistency and accuracy of the solution procedure,   the
time-dependent evolutions of non-rotating heat bubbles during their rise to the surface of
a white dwarf are followed and compared with previous calculations.
}
{In the rotating case and depending on the viscosity parameter, $\alpha_\mm{tur}$, it is found that the
viscously-initiated  fronts at the center of bursts propagate at much faster speed than the
fluid motion.
These fast fronts act to decouple angular momentum from
matter: angular momentum is transported outwards while
matter sinks inwards into the deep gravitational well of the neutron star,
thereby enhancing the compression of matter necessary for initiating ignition,
that subsequently spreads over the whole surface of the neutron star on the viscous time scale.
Based on the numerical simulations, we find that a viscosity parameter
$\alpha_\mm{tur} = \mathcal{O}(0.1)$ is most
suitable for fitting observations of neutron stars during  X-ray bursts. It is argued that the
spin up observed in the cooling tails of X-ray bursts is a transient phase, which eventually should be
followed by a spin down phase. This delay can be attributed to a significant lengthening
of the viscous time scale due to rapid cooling of matter in the outer layers.}

   {}

\maketitle
%

\date{Received ... / Accepted ...}

 \section{Introduction}

Part of the observed neutron stars (NSs) belong to the family of ultra-compact objects, in which general relativistic
effects are prominent \citep{Shapiro1983, Stergioulas2003, Psalti2008}. Depending on the equation of state, NSs may live even
inside their last stable orbits, making the conversion efficiency of gravitational energy
into radiation even larger than that of accreting Schwarzschild black holes
\citep[BHs,][]{Camenzind2007}.
In low mass X-ray binaries, relativistic jets have been also observed to emanate from around
accreting NSs with bulk Lorentz factors that are comparable to those
emanating from around black holes \citep{Migliari2006, Migliari2008}. \\
On the simulation site, numerous general relativistic hydrodynamical calculations
have been carried out to study different aspects of NSs, such as formation, merger, inner structure,
accretion or jets around NSs \citep[][see also the references therein]{Thielemann1990, Liebendorfer2002, Marti2003, ozel2003, Shibata2003, McKinney2006, Shibata2006, Abdikamalov2008, Zachariah2008}

\cite{Duez2004}  carried out general relativistic calculations to study the formation of
hypermassive NSs, taking into account the effect of viscosity. The authors found that
viscosity drives the NS's inner core into rigid rotation and simultaneously
transports angular momentum outwards into the outer envelope.
As a consequence, the core is found to contract in a quasi-stationary
manner while the outer layers expand to form a differentially rotating torus.

This behavior is similar to accretion of matter via rotating disks. Here the
 viscosity acts to decouple matter from angular momentum, in that it transports
 angular momentum outwards, while forcing the matter to sink deeper into the gravitational well
 of the central object\citep{Pringle1981}.
 In the absence of viscosity,
  angular momentum as well as magnetic fields in ideal MHD  are frozen-in
 to the matter. Thus, while strong magnetic fields are essential to enable rigid rotation,
  viscosity, on the other hand,  drives the outer layers into differential rotation.

The type of rotation in the outer layers, differential or rigid rotation, may have
profound effects on the conditions leading to the  X-ray bursts observed on NSs \citep{Bildsten1999, Spitkovsky2002}.

Indeed, recent observations of X-ray bursts revealed the so called burst oscillations, in which
a spin-up or spin-down of the NSs in their cooling tails have been detected,
reaching a plateau on the asymptotic limit \citep[see][ for a detailed discussion]{Strohmayer1997, Strohmayer1999, Strohmayer2001}. It has been argued that the increase/decrease of the spin of NSs during bursts
is connected to the redistribution of angular momentum of the thermonuclear shell
\citep{Strohmayer2001b}.
Accordingly, when a thermonuclear shell starts to expand at the burst
onset, the moment
of inertia increases while its spin decreases. When the shell starts to contract and subsequently
recouples to the NS, the inertia decreases and the spin increases.

Also, several X-ray bursts on NSs display a spin down rather than spin up in their cooling tails
\citep{Strohmayer1999c}.
In this case, however, it was suggested that the spin down
 probably begins anew episode of thermonuclear energy release, most likely
in layers underlying those responsible for the initial runaway.

\cite{Cumming2003}, however, investigated in detail the hydrostatic expansion during bursts
 and the expected change of spin due to angular momentum conservation and concluded, that
 a shell expansion/contraction alone cannot explain the mechanisms underlying the observed
 spin-up/down of the NSs during bursts.
 Noteworthy is that the model in which ignition starts at a point and
 spreads over the whole surface of the NS via burning fronts appears to fit observations,
  which reveals that
 the X-ray emitting area increases during the bursts \citep{Strohmayer1997}.
 However, the role of rotation, the nature of these burning fronts and the manner they affect
 their surrounding are poorly understood.

 In this paper we present a first attempt to model the rotational evolution of thermally induced
  bursts  beneath the atmosphere of a rotating NS and to study the role and effects
  of the viscosity on the redistribution of angular momentum under strong gravitational field
  conditions. Our investigation relies
  on employing a general relativistic hydrodynamical solver, in which turbulent-eddies have
  the effect of friction that gives rise to an enhanced re-distribution
  of angular momentum.

  The paper runs as follows:  In Section 2 we describe the additional viscous operators that have
  been incorporated into the solver to study the viscous-redistribution of angular momentum. The results of several model calculations aimed at studying the distribution of angular momentum during X-ray bursts on NSs are presented and discussed  in Section 3, while in Sec. 4 the results are summarized.
\section{The general relativistic Navier-Stokes equations}
The set of the general relativistic hydrodynamical equations and their derivations
are well described in Sec. 2 of Hujeirat et al. (2008). In this section
we list the viscous operators of the momentum equations, which we have incorporated
into the implicit solver.

The stress energy tensor for viscous flows has the following form \citep{Richardson2002,Font2003, Camenzind2007}:
\beq \rm{ T^{\mu\nu}= \mm{{T_\mm{PF}^{\mu\nu}} +
\left\{\underline{T_\mm{Vis}^{\mu\nu}}\right\}} = \rho~h~u^\mu u^\nu
+ P~g^{\mu\nu} + \left\{
 \underline{
-\eta [ \bar{\sigma}^{\mu\nu} + \DD{\Theta}{3}h^{\mu\nu}]}
 \right\}},
 \label{SET}\eeq
where $\mu,\nu~$ are indices that correspond to the four coordinates $\{t,r,\theta,\varphi\}$ and $\mm{{T_\mm{PF}^{\mu\nu}}, ~T_\mm{Vis}^{\mu\nu}}$ denote the
stress energy tensor due to perfect and viscous flows, respectively.
{\rm P}, $\eta,~\Theta,~$ are the pressure, which is calculated from
the equation of state corresponding to a polytropic or to an ideal
gas, the dynamical viscosity which is assumed to be identical to the
shear viscosity,  and $\rm{\Theta~(\doteq\nabla_\mu u ^\mu)},$ which
measures the divergence or convergence of the fluid world lines,
respectively. $\rm{h^{\mu\nu}= u^\mu u^\nu + g^{\mu\nu}}$ is the spatial
projection tensor, whereas $\bar{\sigma}$ corresponds to the
symmetric spatial shear tensor: \(\rm{ \bar{\sigma}^{\mu\nu} =
\nabla_\varsigma u^\mu h^{\varsigma\nu} + \nabla_\varsigma u^\nu
h^{\varsigma\mu}} .\)  \\
For the X-ray burst calculations, the general relativistic Navier-Stokes
equations are solved using the Boyer-Lindqueit coordinates in the background
 of a slowly rotating NS, with the following metric elements:

\[
g_{\mu\nu} =   \left[
    \barr{cccc}
    g_{tt} & 0 &  0 & g_{t\varphi} \\
    0 & g_{rr} & 0 &  0 \\
    0 & 0 & g_{\thet\thet} & 0  \\
    g_{\varphi t}& 0 & 0 & g_{\varphi\varphi} \earr  \right],
    \]
     \textrm{~~where~~~}
    \beq
   \left\{ \barr{lll}
     g_{tt} &= & \beta_{\varphi}\beta^{\varphi} - \alpha^2 \\
     g_{t\varphi} &=& g_{\varphi t} = \beta_{\varphi} = g_{\varphi\varphi} \beta^{\varphi} \\
     g_{rr} & =& \DD{\bar{\rho}^2}{\Delta}, $~~~~$g_{\thet\thet} = \bar{\rho}^2,$~~~~$ g_{\varphi\varphi}= \bar{\omega}^2 \\
     \Delta & = & r^2 - 2r_gr+ \Omega_{NS}^2 \\
     \bar{\rho}^2& =&  r^2 + \Omega_{NS}^2 \sin^2{\theta}\\
     \Sigma^2 & = & (r^2 + \Omega_{NS}^2)^2 - \Omega_{NS}^2\Delta \cos^2{\thet} \\
     \bar{\omega} & = & \DD{\Sigma}{\bar{\rho}} \cos{\thet} \\
     \alpha^2 & =& \DD{\bar{\rho^2}}{\Sigma^2}{\Delta} \\
     \beta^r & =& \beta^\thet = 0  \\
     \Upsilon & = & \DD{\bar{\rho}^2 \Sigma^2}{\Delta} \cos^2{\thet} \\
    \sqrt{-g} & =& \bar{\rho}^2\cos{\thet} = \alpha
                            \sqrt{\Upsilon}.
                             \earr \right .
\eeq
In this formulation, the parameter ``$\Omega_\mm{NS}$" denotes the spin of the neutron star,
which is taken to be much smaller than the break-up frequency.
$\beta^{\varphi}$ is the frame-dragging frequency associated with the rotation of the NS:
\beq
\rm{\beta^{\varphi} = \Omega_\mm{FD} = [\DD{2G}{c^2}]J_*(\DD{r}{\Sigma^2}) = [\DD{2GM}{c^2}](\DD{1}{r^3})R^2_{NS}\Omega_{NS}},
\eeq
 where $\rm{J_* = I_* \Omega_\mm{NS}},$  and $\rm{I_* = \DD{2}{5} M_{NS} R^2_{NS}}$ is the moment of inertia of the NS. The parameters:
$\rm{c,\, M_\mm{NS},\, G,\,r_g (= \DD{GM_\mm{NS}}{c^2}),\, \alpha}$
denote the speed of light,
  mass of the NS, the gravitational constant, the
  gravitational radius and the lapse function, respectively. In writing these expressions,
  we made use of the coordinate transformation $\bar{\thet} = \pi/2
  - \thet$, where we use the latitude $\thet$ instead of the polar distance
  angle $\bar{\thet}$; hence the appearance of "$\cos$" instead of "$\sin$" in
  the metric terms.

  The set of general relativistic Navier-Stokes equations in 3D axi-symmetry can be written as the residual vector equation:
   \beq
   \rm{R = 0.}
   \eeq
   The components of this vector read as follows:


\ben
\item The continuity equation
   \beq
 \rm    R_1 = \DD{\D D}{\D  t} + L1_{r\,\theta} D = 0
   \eeq
\item The radial momentum equation
\beq
\rm R_2= \DD{\D M_r}{\D t} + L1_{r\,\theta} M_r - f_r - L2^r_{r\,\theta} M_r = 0
\eeq
\item The vertical momentum equation
\beq
\rm R_3= \DD{\D M_\theta}{\D t} + L1_{r\,\theta} M_\theta - f_\theta - L2^\theta_{r\,\theta} M_\theta = 0
\eeq
\item The angular momentum equation
\beq
\rm R_4= \DD{\D M_\varphi}{\D t} + L1_{r\,\theta} M_\varphi - f_\varphi - L2^\varphi_{r\,\theta} M_\varphi = 0
\eeq
\item The internal energy equation
 \beq
 \rm{R_5=
 \DD{\D {\cal{E}}^d}{\D t} + L1_{r\,\theta} {\cal{E}}^d
+(\gamma-1)\,{\cal{E}}^d
[\DD{\D u^t}{\D  t} + L1_{r\theta} u^t]
} = 0,
 \label{Eq5}
\eeq
\een
where  $L1_{r\,\theta}$ are first order advection operators that have the form:
\[
\barr{lll}
L1_{r\,\theta} ~q~ &=& \DD{1}{\sqrt{-g}} \DD{\D}{\D r} (\sqrt{-g}~q~ V^r)
               + \DD{1}{\sqrt{-g}} \DD{\D}{\D \thet}(\sqrt{-g}~ q~ V^\thet)\\
              & =& \bar{\nabla}_r \cdot q V^r + \bar{\nabla}_\theta\cdot q V^\theta.
               \earr
\]
$f_{r,\,\theta,\,\varphi}$ are force terms that include  pressure gradients, centrifugal
and gravitational forces acting along the radial, horizontal and azimuthal directions, respectively.
$\mm{D} (\doteq \rho u^t)$ is the modified relativistic mass density.
$\rm{M}_\mu$ are the four-momenta: \(\rm (M_t,M_r,M_\thet,M_\varphi)
\doteq {\DroI}(u_t,u_r,u_\thet,u_\varphi),\) where \( {\DroI} \doteq D
h,\) and $u^t$ is the time-like velocity, \( V^\mu= u^\mu/u^t\) is
the transport velocity and ``h" denotes the enthalpy. \(L2^{\xi}_{r\theta}\) are the spatial
projections of the viscous stress energy tensor
$\mm{T^{\mu\nu}_{Vis}}~$ (see Eq. \ref{SET}) in the respective
direction. These are obtained from the projection of the viscous
tensor along the vector normal to the hyperspace, i.e., constant in
time:
\[
L2^{\xi}_{r\theta} = \nabla_\mu T_\mm{Vis}^{\mu\xi} = \bar{\partial}_\mu
T_\mm{Vis}^{\mu\xi} + \Gamma_{\mu\lambda}^{\xi}
T_\mm{Vis}^{\mu\lambda},
\]
where $\xi=\{r,\theta,\varphi\}$. $\nabla_\mu$ corresponds to the
spatial divergence of a tensor taken in the Boyer-Lindquist
coordinates and $\Gamma_{\mu\lambda}^{\xi}$ are the Christoffel's
symbols of the second kind.\\

For completeness, we re-write the forms of these second order viscous operators explicitly as follows:
\beq
\barr{llll}
\rm{L2^{r}_{r\theta}}
 & = &\rm{\bar{\nabla}_r \cdot \eta }& \rm{[(\DD{\D u^r}{\D r}+\DD{1}{2}(g^{rr}\DD{\D g_{rr}}{\D r}) u^r)~ (u_r u^r +1)} \\
 &   &&\rm{+\DD{\D u_r}{\D r} - \DD{1}{2}(g^{rr}\DD{\D g_{rr}}{\D r}) u^r)~ ((u^r)^2 + g^{rr})} \\
 &   && \rm{-\DD{2}{3} (\bar{\nabla}_r \cdot u^r + \bar{\nabla}_\theta \cdot u^\theta)~(u_r u^r + 1) ] }\\
 & +  & \rm{\bar{\nabla}_\theta \cdot \eta} &\rm{
          [(\DD{\D u^\theta}{\D r}+\DD{1}{2}(g^{\theta\theta}\DD{\D g_{\theta\theta}}{\D r})~ (u_r u^r +1) }\\
&   &&\rm{+\DD{\D u^\theta}{\D \theta} + \DD{1}{2}(g^{\theta\theta}\DD{\D g_{\theta\theta}}{\D \theta})~
           (u_r u^\theta)} \\
&   &&\rm{+\DD{\D u_r}{\D r} - \DD{1}{2}(g^{rr}\DD{\D g_{rr}}{\D r}) u_r)~ (u_r u^\theta)}  \\
&   &&\rm{+\DD{\D u_r}{\D \theta} - \DD{1}{2}(g^{rr}\DD{\D g_{rr}}{\D \theta}) u_r)~ ((u^\theta)^2 +
                g^{\theta\theta})} \\
&   && \rm{-\DD{2}{3} (\bar{\nabla}_r \cdot u^r + \bar{\nabla}_\theta \cdot u^\theta)~(u_r u^r + 1) ]},
\earr
\eeq

\beq
\barr{llll}
\rm{L2^{\theta}_{r\theta}}
 & = &\rm{\bar{\nabla}_r \cdot \eta }& \rm{[(\DD{\D u_\theta}{\D r}-\DD{1}{2}(g^{\theta\theta}\DD{\D g_{\theta\theta}}{\D r}) u_\theta)~ (u^r)^2 + g^{rr}) }\\
 &   && \rm{-\DD{2}{3} (\bar{\nabla}_r \cdot u^r + \bar{\nabla}_\theta \cdot u^\theta)~(u_\theta u^r) ] }\\
 & +  & \rm{\bar{\nabla}_\theta \cdot \eta} &\rm{
          [(\DD{\D u^\theta}{\D r}+\DD{1}{2}(g^{\theta\theta}\DD{\D g_{\theta\theta}}{\D \theta})~
           (u_\theta u^\theta +1) }\\
 &&& \rm{-\DD{2}{3} (\bar{\nabla}_r \cdot u^r + \bar{\nabla}_\theta \cdot u^\theta)~(u_\theta u^\theta + 1)] }\\
 &&& \rm{+ (\DD{\D u^\theta}{\D \theta}-\DD{1}{2}(g^{\theta\theta}\DD{\D g_{\theta\theta}}{\D \theta})
           u^\theta)~(u^\theta)^2 + g^{\theta\theta})]},
\earr
\eeq

\beq
\barr{llll}
\rm{L2^{\varphi}_{r\theta}}
 & = &\rm{\bar{\nabla}_r \cdot \eta} &\rm{
    [(\DD{\D u_\varphi}{\D r}-\DD{1}{2}(g^{\varphi t}\DD{\D g_{\varphi t}}{\D r}
                                        + g^{\varphi \varphi}\DD{\D g_{\varphi \varphi}}{\D r}
                                         ) u_\varphi)~ (u^r)^2 + g^{rr})} \\
 &   &&\rm{ + (\DD{\D u_\varphi}{\D \theta}-\DD{1}{2}(g^{\varphi t}\DD{\D g_{\varphi t}}{\D \theta}
                                        + g^{\varphi \varphi}\DD{\D g_{\varphi \varphi}}{\D \theta}
                                         ) u_\varphi)~ (u^r u^\theta) }\\
 &   &&\rm{ - (\DD{1}{2}(g^{rr}\DD{\D g_{\varphi t}}{\D r})~u^\varphi u_\varphi u^t
        - (\DD{1}{2}(g^{rr}\DD{\D g_{\varphi \varphi}}{\D r})~u^\varphi (u_\varphi u^\varphi +1)]}\\
 & +  &\rm{ \bar{\nabla}_\theta \cdot \eta} & \rm{
          [(\DD{\D u_\varphi}{\D r} -\DD{1}{2}(g^{\varphi t}\DD{\D g_{\varphi t}}{\D r}
                                        + g^{\varphi \varphi}\DD{\D g_{\varphi \varphi}}{\D r})~
                                         (u^r u^\theta)} \\
 &   && \rm{+ (\DD{\D u_\varphi}{\D \theta}-\DD{1}{2}(g^{\varphi t}\DD{\D g_{\varphi t}}{\D \theta}
                                        + g^{\varphi \varphi}\DD{\D g_{\varphi \varphi}}{\D \theta}
                                         )u_\varphi)~ ((u^\theta)^2 + g^{\theta\theta}}\\
 &   && \rm{        -(\DD{1}{2}g^{\theta\theta}\DD{\D g_{\varphi t}}{\D \theta})~ u^\varphi u_\varphi u^t
          -(\DD{1}{2}g^{\theta\theta}\DD{\D g_{\varphi \varphi}}{\D \theta})~ u^\varphi (u_\varphi u^\varphi +1) ] }, \\
\earr
\eeq
where $g_{\mu\nu}$ is the contravariant form of the metric tensor $g^{\mu\nu}.$

\subsection{Simplifying considerations}
Most of the above-described collection of viscous terms contains highly non-linear, first  and
second order operators, some of which are difficult to handle numerically or unnecessarily enlarge
the band-width of the coefficient matrix, while others may decelerate, rather than accelerate,
the convergence of the numerical solution procedure. In 3D axi-symmetry, few of
these terms can be simplified or even neglected without violating the physical consistency of
the numerical scheme.

\begin{figure}[htb]
\centering {\hspace*{-0.2cm}
\includegraphics*[width=5.5cm, bb=35 510 270 735,clip]
{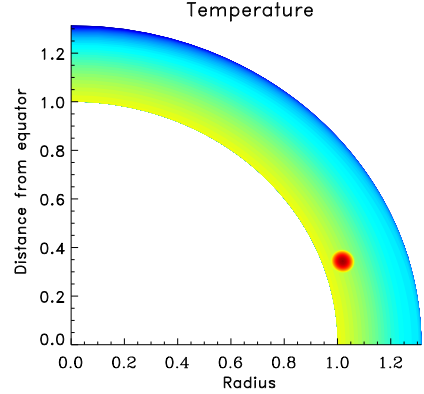} 
 }
\caption{\small The starting distribution of the temperature in the domain of calculation. Low (high) values
 correspond to blue (yellow) color. The red color corresponds to the hot bubble.} \label{Fig1}
\end{figure}

To outline our simplification strategy, we first mention the following relevant issues:

\ben
\item In most astrophysical fluid flows the molecular viscosity is too small to be
       relevant on observationally reasonable time scales. Thus, in the absence of other forms of
       viscosity, the
      above operators can be safely neglected.\\
      Moreover, these operators must vanish asymptotically whenever the fluid velocity approaches
      the speed of light.
\item Turbulent viscosity is more common in  modeling astrophysical fluid dynamics.
      In rotating astrophysical flows,  turbulent viscosity is a fundamentally important
      mechanism for angular momentum transport.
      Therefore, the viscous operators of the angular momentum equation are important
      and should converge to the usual Newtonian form whenever the velocity becomes sub-relativistic.

\item The viscous operators appearing in the radial and vertical momentum equations act, in general,
       to diffuse and smooth strong
      velocity-gradients. The mixed-derivative appearing in these two equations act to mainly enhance the
      viscous-coupling between the velocity components.
\een

Therefore, our simplification strategy relies on considering only second
      order, mixed-free and Laplace-like operators. In numerical matrix algebra, such operators
      generally enhance
      the diagonal dominance of the coefficient matrix and stabilize its inversion procedure.
\begin{figure*}[htb]
\centering {\hspace*{-0.2cm}
\includegraphics*[width=9.65cm, bb=35 47 357 738,clip]
{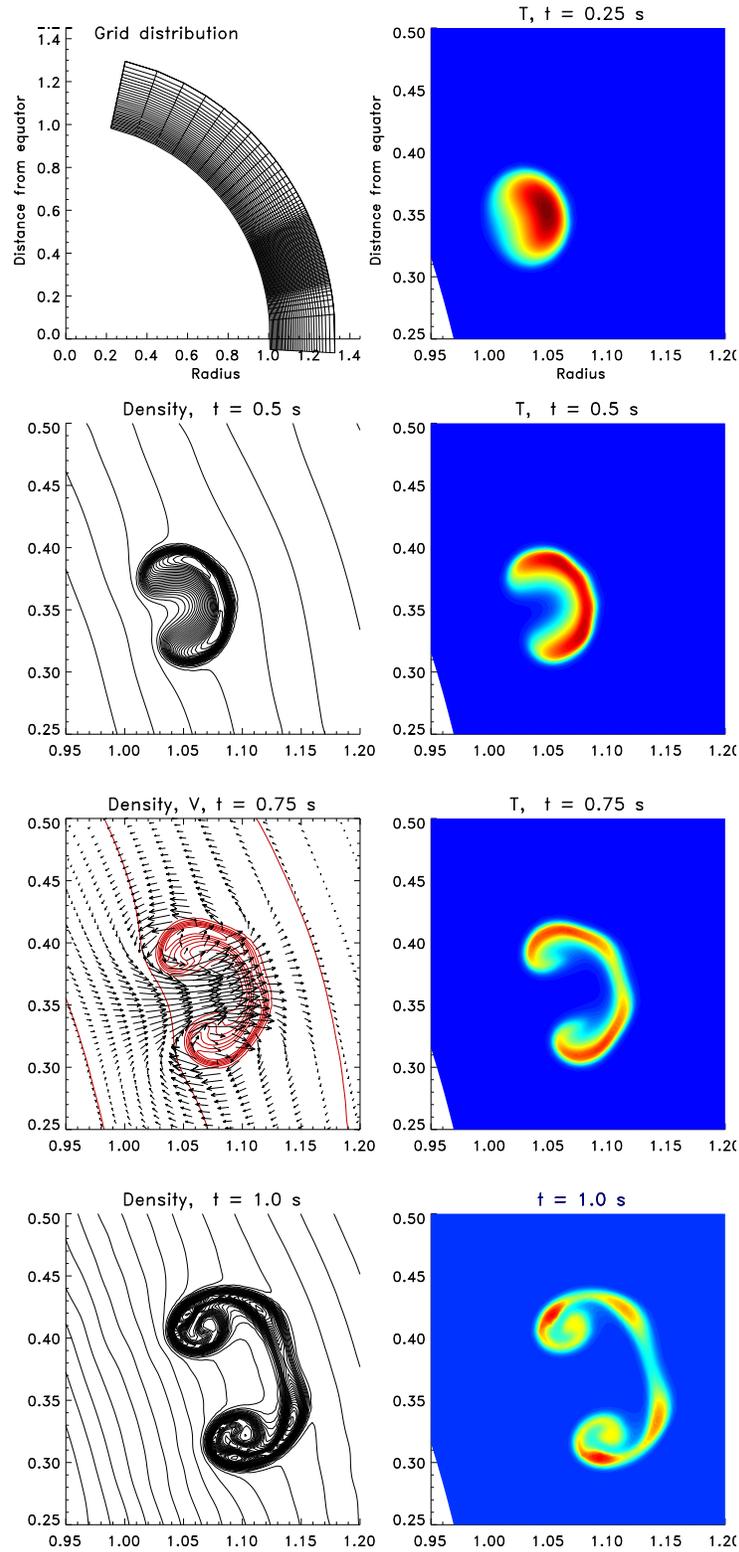} 
}
\caption{\small a rising bubble in the white dwarf atmosphere. The domain of calculation
is covered by $100\times170$ non-uniformly distributed finite volume cells (top left panel).
The colored images are snapshots of the temperature of the rising bubble after 0.25 , 0.5.
0.75 and 1.0 second after the flash. low-to-high values of temperature correspond to
blue-to-red. On the left panel, snapshot of the density distribution at 0.5, 0.75 and 1.0 seconds
in black-white are shown.} \label{Fig2}
\end{figure*}

 Specifically, using our time-implicit formulation, the following operators have been included into
  the numerical solver:
\beq
\barr{llll}
\rm{\tilde{L2}^{r}_{r\theta}}
 & = &\rm{\bar{\nabla}_r \cdot \{\eta }&\rm{
  [2\DD{\D u^r}{\D r}  -\DD{2}{3} (\bar{\nabla}_r \cdot u^r )]\}~(u_r u^r + 1)}  \\
 & +  & \rm{\bar{\nabla}_\theta \cdot \{\eta} &\rm{
           (\DD{\D u_r}{\D \theta})\}~ ((u^\theta)^2 + g^{\theta\theta})} \\
\earr
\eeq

\beq
\barr{llll}
\rm{\tilde{L2}^{\theta}_{r\theta}}
 & = &\rm{\bar{\nabla}_r \cdot \{\eta} & \rm{[(\DD{\D u_\theta}{\D r})~ (u^r)^2 + g^{rr})} \\
 & + &\rm{ \bar{\nabla}_\theta \cdot \{\eta} &\rm{
  (\DD{\D u^\theta}{\D \theta})-\DD{2}{3} \bar{\nabla}_\theta \cdot u^\theta)~(u_\theta u^\theta + 1)
  + (\DD{\D u^\theta}{\D \theta})~(u^\theta)^2 + g^{\theta\theta})]\}}
\earr
\eeq

\beq
\barr{llll}
\rm{\tilde{L2}^{\varphi}_{r\theta}}
 & = &\rm{\bar{\nabla}_r \cdot \eta} &\rm{
    [(\DD{\D u_\varphi}{\D r}-\DD{1}{2}(g^{\varphi t}\DD{\D g_{\varphi t}}{\D r}
                                        + g^{\varphi \varphi}\DD{\D g_{\varphi \varphi}}{\D r}
                                         ) u_\varphi)~ (u^r)^2 + g^{rr}) }\\
 &   && \rm{- (\DD{1}{2}(g^{rr}\DD{\D g_{\varphi t}}{\D r})~u^\varphi u_\varphi u^t
        - (\DD{1}{2}(g^{rr}\DD{\D g_{\varphi \varphi}}{\D r})~u^\varphi (u_\varphi u^\varphi +1)]}\\
 & +  & \rm{\bar{\nabla}_\theta \cdot \eta} &\rm{
  (\DD{\D u_\varphi}{\D \theta}-\DD{1}{2}(g^{\varphi t}\DD{\D g_{\varphi t}}{\D \theta}
                                        + g^{\varphi \varphi}\DD{\D g_{\varphi \varphi}}{\D \theta}
                                         )u_\varphi)~ ((u^\theta)^2 + g^{\theta\theta}) }\\
 &   && \rm{-(\DD{1}{2}g^{\theta\theta}\DD{\D g_{\varphi t}}{\D \theta})~ u^\varphi u_\varphi u^t
          -(\DD{1}{2}g^{\theta\theta}\DD{\D g_{\varphi \varphi}}{\D \theta})~ u^\varphi (u_\varphi u^\varphi +1) ] } \\
\earr
\eeq
It can be easily verified that in the non-relativistic regime the radial component of the
diffusion operator $\rm{\tilde{L2}^{\varphi}_\mm{r\varphi}}$ reduces to the classical Newtonian form:
\beq
\rm{\tilde{L2}^{\varphi}_{r} = \DD{1}{r^2}\DD{\D}{\D r} r^4 \eta \DD{\D \Omega}{\D r}},
\eeq
where $\eta = \rho \nu$ and $\nu$ denotes the kinematic viscosity.

\begin{figure*}
\centering {\hspace*{-0.2cm}
\includegraphics*[width=10.5cm, height=20cm, bb=34 42 352 740,clip]
{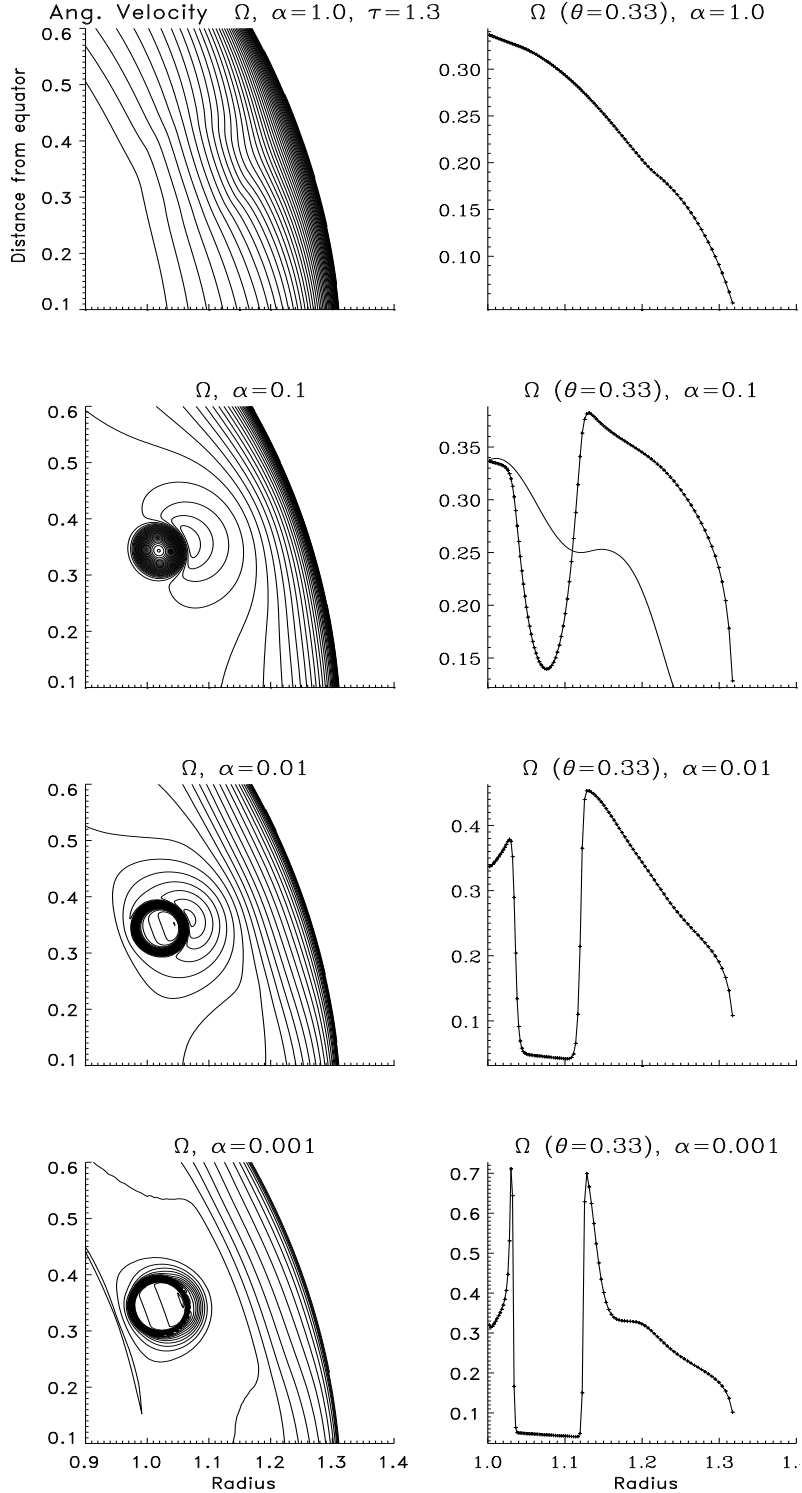} 
}
\caption{\small The distribution of the angular velocity in the bursting region
 for four different viscosity parameters
$\alpha_\mm{tur}$ shortly after burst events $(\approx 0.1 ms)$. In the left panel 30 uniformly
distributed isolines of the angular frequency $\Omega$ are shown.
On the right panel we display the radial profiles of $\Omega$ across the bursting region.
The solid line in ``(b2)´´ corresponds to a relaxed $\Omega-$profile after 10 ms.}
\label{Fig3}
\end{figure*}

\subsection{Viscosity prescription}
 Similar to classical accretion disks, we assume that molecular viscosity 
 is too small to have a significant
  effect on the angular momentum distribution on a short time scale such 
  as the thermonuclear one.
 
  Therefore, we adopt the turbulent viscosity prescription:
  \beq
  \nu_\mm{tur} \doteq <V_\mm{tur}><\ell_\mm{tur}>\approx \alpha_\mm{tur} 
  V_\mm{S} \times \alpha_\mm{2} R_\mm{NS},
  \eeq
where $<V_\mm{tur}>,~~ <\ell_\mm{tur}>$ correspond to mean values of velocity 
and length scale
of eddies in a turbulent medium, respectively. 
 \clearpage
These are set to be respectively 
smaller than the sound speed $V_\mm{S}$ and smaller than the
radius of the NS. Thus, $\alpha_\mm{tur},~\alpha_\mm{2}$ are constants that are
set to be smaller than unity.   In the present paper, all model calculations assume
$<\ell_\mm{tur}> = 0.1 R_\mm{NS} $, i.e., $\alpha_\mm{2}= 0.1$, whereas the parameter
$\alpha_\mm{tur}$ may differ from one model
calculation to another.
\section{Heat bubble calculations}
\subsection{Numerical solution method}
 The set of hydrodynamical equations are solved using a pre-conditioned
 defect-correction iteration procedure. The matrix equation to be solved in each iteration is:
 \beq
 \mm{\tilde{A}}\mu = d,
 \eeq
 where $\tilde{A}$ is a preconditioner, i.e., a coefficient matrix that is similar to the
 Jacobian J, but which is much easier to invert. J is obtained by calculating the entries resulting from
 $\D R/\D q,$ where R  denotes the vector of equations (Eq. 4) and q the vector of variables. \\
 In this formulation $\mu = q^{i+1}-q^{i}$ corresponds to the correction between two successive
 iterations and d is the
 defect \citep[for a detailed description of the method see][]{ Hujeirat2005, Hujeirat2008}.

 We mention that the solver employed here relies on the conservative
 formulation of the hydrodynamical equations, using the finite volume formulation.
 For strongly time-dependent simulations, an advection scheme of third order spatial
 and second order temporal accuracies is used \citep{Hujeirat2005}.
 As pre-conditioner we use the approximate factorization method (AFM), which is proven
 to be most appropriate for modeling weakly and strongly low Mach number flows \citep{Hujeirat2007}.
\begin{figure}[htb]
\centering {\hspace*{-0.2cm}
\includegraphics*[width=5.5cm, bb=50 277 281 504,clip]
{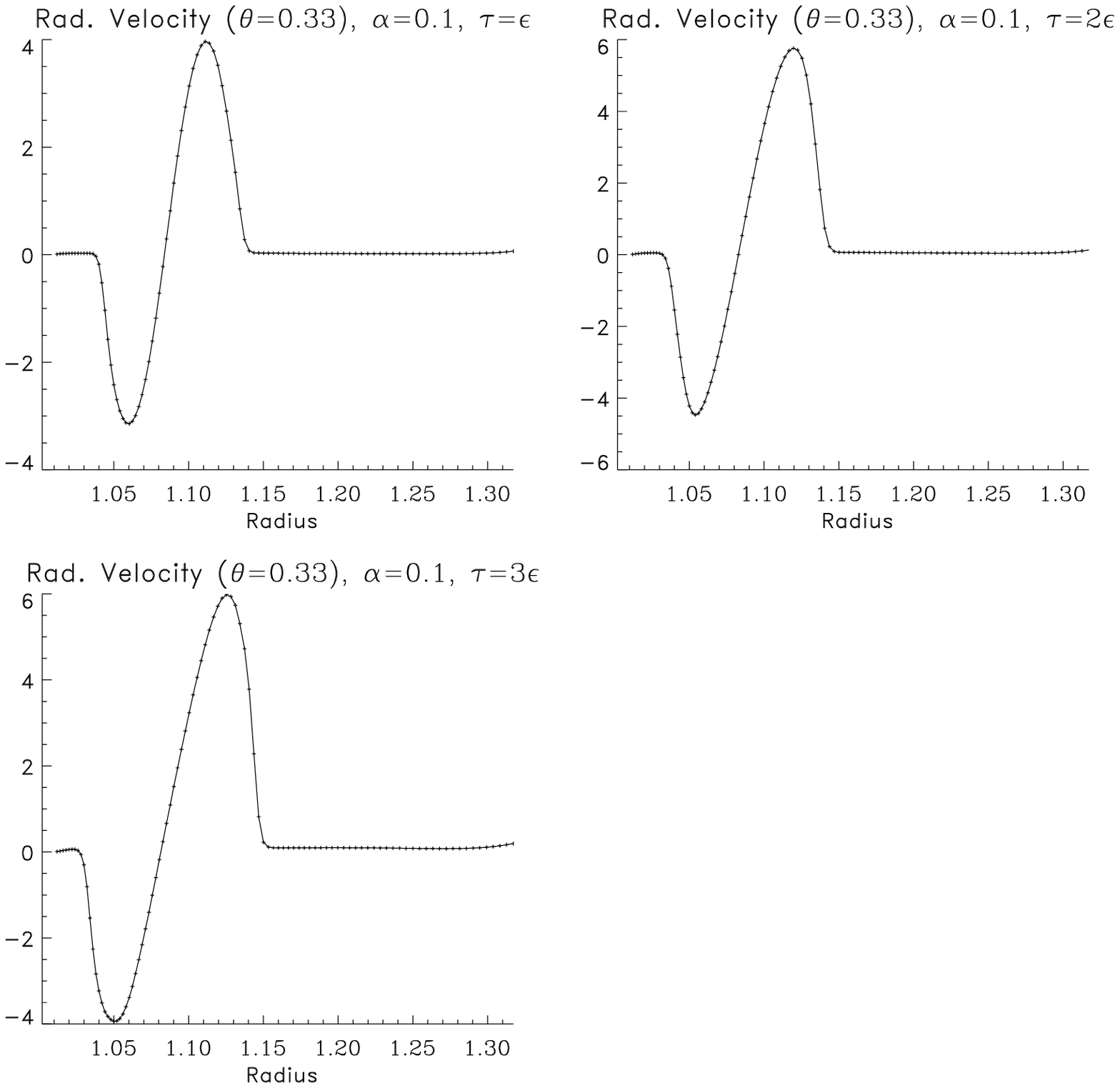} 
}
\caption{\small The profile of the radial velocity across the heat bubble shortly after
the burst (0.3 ms), using   $\alpha_\mm{tur}=0.1$. Obviously, the outward-oriented
fluid motion is much stronger than the inwards one.}
\label{Fig4}
\end{figure}

\subsection{Heat bubble propagation in the atmosphere of a non-rotating white dwarf}
Rising bubbles in stellar environments have been extensively studied by
Almgren at al. (2006, see also the references therein).

To test the capability of our solver at capturing the propagation of strongly time-dependent heat bubbles
in strong gravitational fields, such as X-ray bursts on neutron stars, we adopt the same setup of the heat bubble problem as described
 in Sec. 4.3 of \citet{Almgren2006}.

The domain of calculation is restricted to the first quadrant:
\[
\barr{lll}
\mathcal{D} &=& [R_\mm{in}\leq r \leq R_\mm{out}]\times[0 \leq \theta \leq \pi/2]\\
            &=& [1 \leq r \leq 1.35 ]\times[0 \leq \theta \leq \pi/2],
            \earr
\]

where length scales are measured in units of the radius of the star's core.\\
The domain $\mathcal{D}$ is divided into non-uniformly distributed finite
volume cells: 100 in the radial and 170 in the horizontal direction, where the minimum grid spacings
is set to coincide with the center of the initial heat bubble.

 The initial conditions are constructed by carrying out 3D axi-symmetric time-dependent calculation
  to obtain the hydrostatic equilibrium. A small region is then chosen (Fig. \ref{Fig1}), where the matter is replaced
  by a thinner but much hotter plasma, while keeping it in pressure equilibrium with the surrounding media.

  In Fig. \ref{Fig2} we show several snapshots of the rising bubble. Obviously, these results
  agree quite well with the unsplit and the low Mach number schemes presented by
  \cite{Almgren2006}.

\begin{figure*}[htb]
\centering {\hspace*{-0.2cm}
\includegraphics*[width=17.5cm, bb=35 510 518 735,clip]
{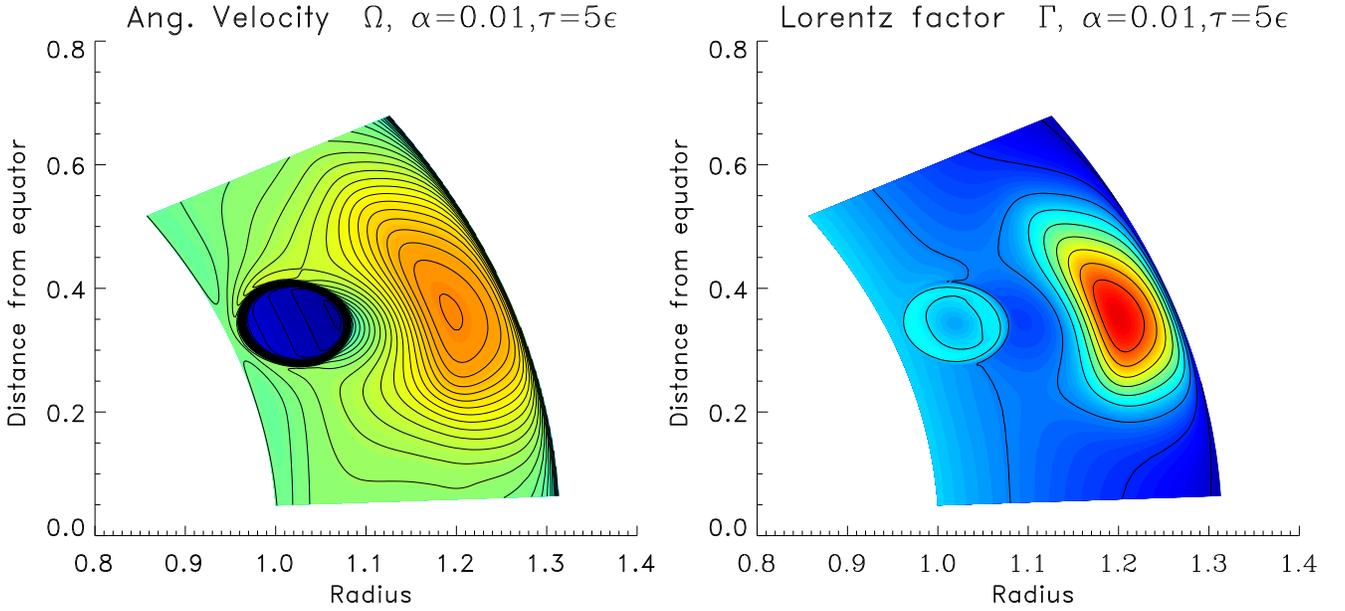}  
}
\caption{\small 25 isolines of the angular velocity (left) and of the Lorentz
factor (right panel) overlayed on their corresponding colored images after 0.5 ms are shown,
using $\alpha_\mm{tur}=0.01.$
Blue-to-red color correspond to low-to-high values of $\Omega$ an $\Gamma$.}
\label{Fig5}
\end{figure*}

\subsection{Rotating heat bubbles in deep gravitational fields of neutron stars }

Similar to the Sec. (3.2), we apply our general relativistic solver to model
the rise of a rotating bubble starting from below the atmosphere of a rotating neutron star.
The radius of the neutron star's core $R_\mm{NS}$ is set to be equal to the following radius,
\beq
R_\mm{NS} = 3\, r_\mm{g}\,(1 + \sqrt{1-\Omega^2_\mm{NS}}),
\eeq
which is smaller than the classical last stable orbit of a Schwarzschild black hole.
 The inner boundary of the domain of calculation is taken to be the radius of the core, whereas
 the outer boundary is located at  $R_\mm{out} =1.35 R_\mm{NS}.$
 The core is set to rotate at 0.4 the break-up velocity, whereas
 $\Omega$ at the outer radius is 10 times smaller.
 The matter in the domain is set to adjust its rotation to the boundaries through viscous interaction.

 The initial distribution of the variables are obtained by solving the general relativistic Navier-Stokes
 equations to obtain stationary and differentially rotating flow configurations.
 As in the previous section, the heat bubble
 is injected into the domain through replacing the medium at a certain location by a hot and tenuous plasma, while
 keeping it in pressure and rotational equilibrium with the surrounding media.

 Unlike the calculations in the pervious section, the purpose of the present calculations is to
 unveil the response of the surrounding region to violent events associated with a dramatic
 change in the distribution of the angular momentum.

 Therefore we run several numerical calculations with $\alpha_\mm{tur}~=~1.0,~0.1,~0.01,~0.001.$
 The initial stationary configurations have been obtained using the corresponding
 value of $\alpha_\mm{tur}$.
  The results are shown in Fig. \ref{Fig3} and can be summarized as follows:
  \ben
  \item All model calculations show a pronounced deficiency of angular momentum at the
        central part of the burst,  accompanied with a significant increase  at the boundary of
        the bubble.
        Thus, the burst leads to the formation of a dynamically  unstable flow-configuration:
        a shell of slow rotating matter  is bounded both from below and from above
        by relatively fast rotating matter.

  \item The viscous-induced fronts of angular momentum are found to propagate outwards
        at much faster speed than inwards. This implies that the matter in the deeper layers
        adapts its conditions to the inner boundary much faster than at the outer boundary. This
        is a consequence of the adopted prescription of the viscosity $\eta,$ which is more effective
         in hotter and denser regions of the plasma.
        Also, the outwards-oriented velocity is obviously larger than the inward one (Fig. \ref{Fig4}).
  \item The difference between the rotational velocity at the center of the bubble and that at its boundaries
          becomes more significant, the smaller the $\alpha_\mm{tur}-$parameter is chosen.
  \een

  In the $\alpha_\mm{tur}=1.0$ case (see Fig. \ref{Fig3} (a1)), the rise time of the bubble is found to be extremely
  long compared to those obtained with smaller $\alpha_\mm{tur}.$

  Here the viscous pressure in the radial direction $P^r_\mm{visc}\sim\alpha_\mm{tur}\DD{\D u }{\D r}$
  has an opposite sign and acts to reduce the effects of the thermal pressure $P_\mm{th}$.
  In the extreme case, when  $P_\mm{th} + P_\mm{visc} \rightarrow 0,$  the effective sound speed,
  $V^{eff}_\mm{S} \sim \delta(P_\mm{the} + P_\mm{visc})/\delta \rho \rightarrow 0,$ hence the
  propagation time $\tau_\mm{pro}\sim R/V^{eff}_\mm{S} \rightarrow \infty .$

  On the other hand, $\tau_\mm{pro}$ becomes of the order of one second, when using
  $\alpha_\mm{tur} \sim \mathcal{O}(10^{-1})$, which fits well into the observed duration of burst events
  on NSs.

  In the low $\alpha_\mm{tur}$ cases, (see Fig. \ref{Fig3} b1,c1,d1 and Fig. \ref{Fig5}), the significant increase of the
  rotational velocity at the outer $\Omega-$fronts is obvious, but unreasonably large.
  In the Fig. \ref{Fig2} (d1), the matter at the outer front is found to be gravitationally unbound
  to the central NS, giving rise to strong outflows. \\
  However, as outflows during X-ray bursts can be excluded on observational grounds, we conclude
  that $\alpha_\mm{tur}$  must acquire much larger values, and that specifically $\alpha_\mm{tur} \sim \mathcal{O}(10^{-1})$.

\begin{figure}
\centering {\hspace*{-0.2cm}
\includegraphics*[width=6.5cm, bb=14 14 510 335,clip]
    {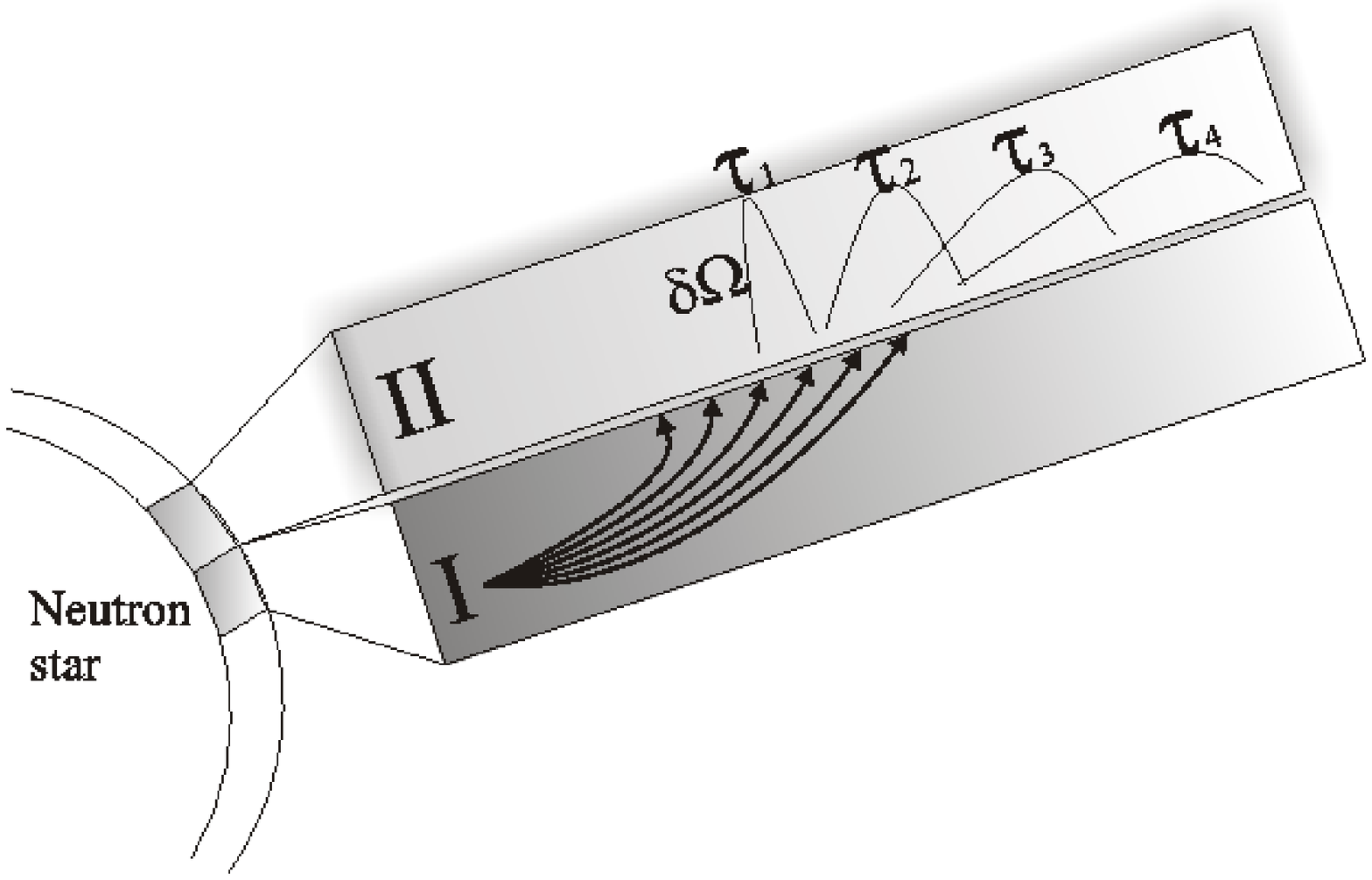} \\ 
\includegraphics*[width=6.5cm, bb=277 512 522 737,clip]
{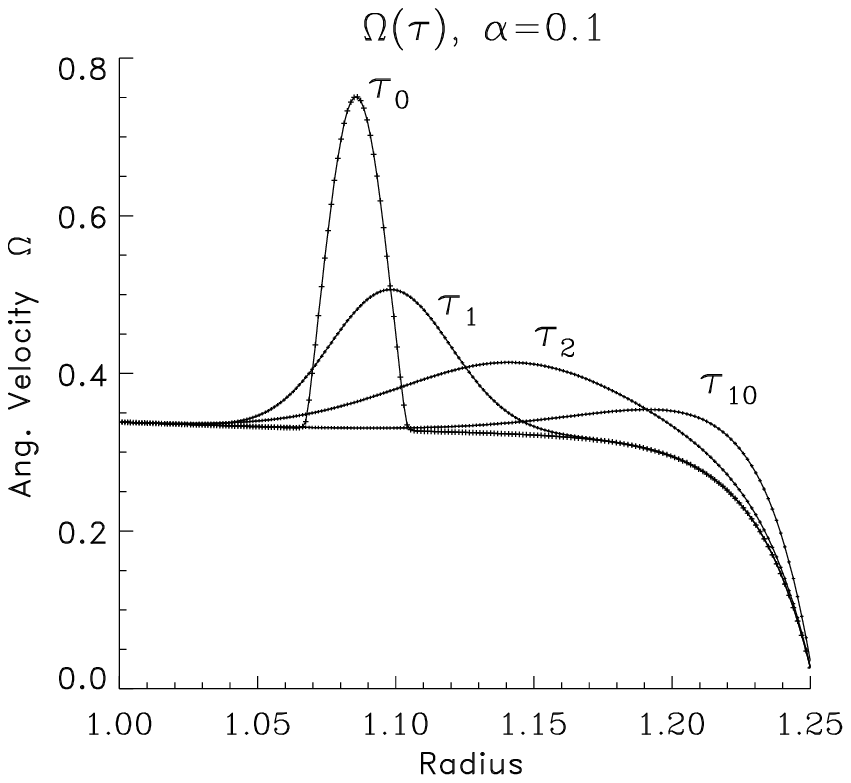}
}
\caption{\small Successive snapshots of the omega-profiles in the outer layers of a
neutron star. Shell I shows the burst-induced propagation of Omega-fronts into the
neighboring shells. In shell II, the viscous-induced reaction to the Omega-fronts is shown.
This schematic picture  shows  that
viscosity transport the excess of angular momentum preferably in the outward direction.
The lower figure, which is obtained  using hydrodynamical calculations,
clearly confirms this behavior. } \label{Fig6}
\end{figure}

  To study the viscous-reaction of the matter in the adjusting layers to the sudden increase of
   rotation induced by a burst, we have run separate
  calculations in the following manner. A solution for the hydrodynamical equations including
   rotation has been hydrodynamically  calculated. As a second step, we modified the $\Omega-$profile by
  including a Gaussian perturbation of the form depicted in Fig.
  \ref{Fig6}/$\tau_\mm{0}$. We then followed  the time evolution of this profile on the viscous time scale. The profiles $\tau_\mm{1}, \tau_\mm{2} , \tau_\mm{10}$ correspond to $\tau_\mm{visc}/10,\tau_\mm{visc}/5,\tau_\mm{visc}.$\\
  These calculations show that the effect of turbulent viscosity is to mainly transport angular momentum
  outwards. As a consequence, the deficiency in the rotational support in the deep layers enhances
   the compression of matter and give rise
  to an additional burst in the neighboring shell. This chain of reactions may run away to spread over the whole surface of the
  NS on the viscous time scale, which is of the order of one second, assuming $\alpha_\mm{tur} \approx 0.1.$

  When the outer layers cool, the turbulent viscosity decreases and the corresponding viscous time scale
  increases as $\tau_\mm{visc} \sim 1/\surd{T}$. This implies that after the burst, the time scale needed for the outer layers to adjust their rotation to the bulk of the star might lengthen by an additional
  order of magnitude.
  As a consequence, the observed spin up of NSs in their  cooling tail is a manifestation of the increased
  rotational velocity of the outer layers caused by the burst events, but which, eventually, should decrease at later times.

\section{Summary \& Conclusions}

In this paper we have presented the set of general relativistic Navier-Stokes
equations in 3D axi-symmetry, using the Boyer-Lindquist coordinates in the background of a slowly
rotating neutron star.
We have shown that, for modeling relativistic viscous flows, the collection of viscous operators can be reduced considerably into a smaller class, which
consists of the dominant second order operators, and which subsequently enhances
convergence of the implicit solution procedure.\\
The set of equations are solved numerically using an implicit solution procedure, which is
based on a pre-conditioned defect-correction iterative method. Similar to
Taylor-flows between concentric spheres, we use the "Approximate Factorization Method"
as a pre-conditioner, which has superior converging properties over other non-symmetric methods, such as
the black-white line Gauss-Seidel method.
In the non-rotating case, we have shown that the solver is capable of reproducing the
time-evolution of heat bubbles during their rise to the surface of the white dwarfs,
as reported by \citet{Almgren2006, Nonaka2008}, accurately.

In the rotating case, it has been shown that viscous-generated fronts inside heat bubbles
propagate into the surrounding quite rapidly. The effect of these front is mainly to transport angular momentum
 to the outer layers, leaving the matter in the deeper layer with less rotational support,
 hence more compressed.\\
 The increased  rotational velocity of the outer layers may be connected to
  the observed spin up of neutron stars during X-ray bursts. However, this increase will
  eventually be followed by a spin down on later times when the outer layers cooled down
  to their pre-burst thermal state.\\
  Our numerical results show that a viscosity parameter of the order of $\alpha_\mm{tur}\sim 0.1$
  is the most suitable value for fitting  observations of NSs during X-ray bursts. A much larger value
  yields propagational time that is much larger than one second, whereas smaller values yield unstable
  shell configurations and gives rise to gravitationally unbound outflows.

 In addition, a possible mechanism that may underly the rapid spreading of
 burning ignition fronts has been presented. Accordingly, the viscous-generated
 fronts inside a heat bubble may transport angular momentum into the horizontally
 adjusting layers. The viscosity then acts to decouple matter from angular
 momentum, subsequently enhancing compression in the deeper layers and giving rise
 to thermonuclear runaway. This chain of reactions may run away to spread over the whole
 surface of the NSs on the viscous time scale
 $\tau_\mm{vis}\sim 1\, \rm{s}$ .

Concerning frame dragging, while included in the present simulations,
its effects are still to be quantified. Also, the effects of magnetic fields and
thermal diffusion is the subject of an ongoing work.

{\bf Acknowledgment} AH thanks Max Camenzind for the 
the valuable discussions and Bernhard Keil and Paul P. Hilscher for carefully
reading the manuscript.
      This work is supported by the Klaus-Tschira Stiftung under the
project number 00.099.2006.


\end{document}